\documentclass[aps,pra,twocolumn]{revtex4}
\usepackage{epsfig}
\usepackage{amssymb}
\usepackage{graphicx}

\begin{document}


\title{Quantum description and properties of electrons emitted from pulsed nanotip electron sources}

\author{$^{1}$Pavel Lougovski\footnote{email: plougovski2@unl.edu} and $^{1}$Herman Batelaan}

\affiliation{$^{1}$Department of Physics and Astronomy,  University of Nebraska Lincoln, Lincoln, NE 68588, USA }

\begin{abstract}
We present a quantum calculation of the electron degeneracy for electron sources. We explore quantum interference of electrons in the temporal and spatial domain and demonstrate how it can be utilized to characterize a pulsed electron source. We estimate effects of Coulomb repulsion on two-electron interference and show that currently available nano tip pulsed electron sources operate in the regime where the quantum nature of electrons can be made dominant.  
\end{abstract}


\maketitle

\section{Introduction} 
Ever shorter synchronous laser and electron pulses are being developed~\cite{Zewail1, Miller, Krause, Kasevich, Elsaesser} as probes to investigate the spatial and temporal behavior of atoms, molecules and nanostructures. Although attosecond physics is already an exciting and active area of research, no electron pulses shorter than ten femtoseconds that are deliverable on a target of choice have been realized yet. After the initial generation of electrons, the short pulses rapidly disperse due to their energy spread. The energy spread is inherent to the Heisenberg uncertainty principle and pulse compression is thus needed to deliver short pulses on a target. Pulse compression techniques such as RF pill boxes~\cite{Krause} and shaped laser modes~\cite{Zewail1} are being explored for this purpose. 
	
In this paper ultrafast electron dynamics is considered. For one-electron pulses, the pulse duration is limited by the uncertainty principle. This is expected to pose no principle obstacle for current technologies to reach deep into the attosecond regime~\cite{Zewail1}. For multi-electron pulses,  wave packet compression is ultimately limited by Coulomb repulsion and the Pauli Exclusion Principle. To address the question how dynamics of wave packets is influenced by the Exclusion Principle, a theoretical framework is developed in this paper. Using this approach, estimates of the effects of electron degeneracy are made and compared to estimates of the effects of Coulomb repulsion. We show that at the source degeneracies of about $10^{-1}$ are expected in current pulsed electron sources ignoring Coulomb repulsion. The effects of Coulomb repulsion are comparable in magnitude to the dynamical effects of degeneracy. The degeneracy of $10^{-1}$ is three orders of magnitude higher than previously reported for any free electron experiment with a continuous electron source. This is in spite of the fact that continuous field emission tips are considered to be very bright sources and a workhorse for electron microscopy~\cite{SilvermanBook}. Pulsed sources are thus expected to increase the instantaneous brightness and degeneracy considerably. Electron pulse compression may increase the degeneracy even further, and it is expected that the Pauli Exclusion Principle needs to be accounted for in the dynamics of ultrafast free electron pulse studies. 

Our theoretical approach is designed to handle the evolution of electron wave packets that evolve continuously in time from the initial to the final state. Rather then studying the process of electron emission from a nano tip, we assume that a certain multi electron state was emitted by the source. We then are interested in its time evolution, taking into account electron dispersion and Coulomb repulsion. The approach leads to expressions that are related to the intensity correlation functions that one finds for the Hanbury-Brown Twiss effect~\cite{HBT}, but provides a time-dependent picture as well as the usual spatial picture. An example is discussed of the first order electron correlation signal measured with one detector that can be used to determine the magnitude of the electron degeneracy.  That this is possible relies on the consideration of a pulsed source as opposed to the usual continuous source. Continuous sources were used in the historic optical experiment of Hanbury-Brown and Twiss~\cite{HBT}, and in the more recent experiment that demonstrated the same effect for free electrons~\cite{Hasselbach}.

This paper is organized as follows. In Section~\ref{sec:description} the description formalism  is introduced. A model for the density matrix of the source is discussed, and the degeneracy is defined. The values of the degeneracy for a pulsed electron source are calculated in Section~\ref{sec:results} using experimentally feasible parameters. In Section~\ref{sec:interference}  quantum interference effects for two-electron states are discussed. The effects of Coulomb repulsion are estimated in Section~\ref{sec:coulomb}. Summary, conclusions and outlook are provided in Section~\ref{sec:last}.

\section{Description formalism}\label{sec:description}
Electron sources to be considered throughout this paper can be sketched as consisting of a metallic nano tip irradiated by a pulsed laser. The duration of each laser pulse is relatively short -- on the order of 50 femtoseconds. As a result, the tip is exposed to very strong electric fields over a very short period of time. The intensity of the laser can be adjusted and thus the average number of photoelectrons emitted by the tip can be controlled ranging from just a single electron to hundreds of electrons per pulse~\cite{Bat1}. 

We are interested in the scenario when, on average, only a few electrons are emitted per pulse. This would be the regime where the quantum nature of electrons is dominant and where a quantum theory describing the source's statistics is needed.

\subsection{Second Quantization}

The standard formalism of quantum mechanics applies to systems with fixed number of particles. However, this is not the case when one is interested in electron emission sources. The number of electrons emitted by the tungsten nano tip fluctuates from pulse to pulse. In this situation, the second quantization formalism is a natural way to account for the fluctuations within the quantum paradigm. 

Let us briefly review the basics of the second quantized description. Instead of keeping track of individual electrons we will use the notion of electron field. The field is represented by the field operators  $\hat{\Psi}^{\dagger}(\bold{x}),\hat{\Psi}(\bold{x})$ defined continuously in all points of space $\bold{x}$. They obey the following anticommutation relations $\{\hat{\Psi}(\bold{x}),\hat{\Psi}^{\dagger}(\bold{x^{\prime}})\}=\delta(\bold{x}-\bold{x^{\prime}})$ -- a direct consequence of the fermionic nature of electrons. It is also convenient to define creation and annihilation operators $c^{\dagger}_{\bold{k}}$ and $c_{\bold{k}}$ that are connected to the field operators via the Fourier transform: 
\begin{eqnarray}
\hat{\Psi}(\bold{x}) & = & \frac{1}{\sqrt V}\int^{+\infty}_{-\infty}d\bold{k} e^{i \bold{k} \cdot \bold{x}}c_{\bold{k}} \nonumber\\
\hat{\Psi}^{\dagger}(\bold{x}) & = & \frac{1}{\sqrt V}\int^{+\infty}_{-\infty}d\bold{k} e^{-i \bold{k} \cdot \bold{x}}c^{\dagger}_{\bold{k}}, 
\end{eqnarray}
here $V$ is a quantization volume. Like the field operators, the creation and annihilation operators also anticommute i.e.  $\{c_{\bold{k}},c^{\dagger}_{\bold{k^{\prime}}}\}=\delta(\bold{k}-\bold{k^{\prime}})$. 

The next step is to introduce the Hilbert space that encompasses  all possible multi-electron states. It is also known as the Fock space and it is constructed as a direct sum of tensor products of single-particle Hilbert spaces. Among all of the Fock states, the most basic one is the vacuum $|0\rangle$. It corresponds to a situation when no electrons are emitted by the tip. Acting on the vacuum with the creation operator $c^{\dagger}_{\bold{k}}$ will generate a particle (an electron) in a state described by the index $\bold{k}$~\cite{fnote1}. In other words, $c^{\dagger}_{\bold{k}}|0\rangle=|0\cdots0 1_{\bold{k}} 0\cdots0\rangle= |1_{\bold{k}} \rangle $, that is, an electron in the state $|\bold{k}\rangle$ is created from the vacuum. On the other hand, by acting with the annihilation operator $c_{\bold{k}}$ on the one-electron Fock state $|1_{\bold{k}} \rangle $ a particle will be destroyed and the resulting state will be the vacuum again. Analogously, we interpret the action of the field operator  $\hat{\Psi}^{\dagger}(\bold{x})$ upon the vacuum state $|0\rangle$ as a creation of a particle in the position $\bold{x}$, or equivalently an electron in the state $|\bold{x}\rangle$, i.e. $\hat{\Psi}^{\dagger}(\bold{x})|0\rangle =|1_{\bold{x}} \rangle$. Lastly, because the Pauli exclusion principle for electrons imposes that $\{c^{\dagger}_{\bold{k}},c^{\dagger}_{\bold{k^{\prime}}}\}=0$ no two electrons can be found in a one-particle state described by the same state index and thus states like $|2_{\bold{k}}\rangle$ are forbidden. However, the states like $|1_{\bold{k}}\cdots 1_{\bold{k^{\prime}}}\rangle$ are allowed.

\subsection{State Description: Density Matrices}  

Having previously defined electron field operators and the Fock space they act upon we now need to assign a quantum state to our electron source in order to be able to make quantitative predictions about electron statistics.   Let us denote the total number of laser pulses that strike the tip of the source by $M$. Suppose, that after the $i$-th laser pulse a bunch of electrons in the state $|\Psi_{i}\rangle$ is emitted.  Assuming that  $N_{max}$ is the maximal number of electrons that can be emitted per pulse, one can write down $|\Psi_{i}\rangle$ as follows,
\begin{equation}\label{psi_i}
|\Psi_{i}\rangle = \alpha_{0,i}|0\rangle + \alpha_{1,i}|1^{el}\rangle + \cdots + \alpha_{N_{max},i}|N_{max}^{el}\rangle,
\end{equation}
where $|0\rangle$  is the vacuum state (no electrons emmited), $|1^{el}\rangle$ is a state with one electron, $|N_{max}^{el}\rangle$ is a state that contains precisely $N_{max}$ electrons~\cite{fnote2}, and $|\alpha_{j,i}|^2, j=0,\cdots,N_{max}$ is a probability to detect $j$ electrons after the $i$-th pulse.

We are now going to make a major assumption about the nature of our source. We assume that the probabilities to detect $0,1,\cdots,N_{max}$ electrons do not change from pulse to pulse i.e. $|\alpha_{j,i}|$ is the same for all $i$ for a given value of $j$. The assumption can be backed up experimentally. Indeed, at currently achievable laser repetition rates of several MHz, the time interval between two successive electron emission events is several orders of magnitude larger than the time scale of relevant dynamics within the emitted multi-electron states~\cite{Kasevich,Bat2}. In other words, the states $|\Psi_{i}\rangle$ should not be affected by the history of prior electron emission events. Therefore, provided that all experimental parameters relevant to the emission process are stabilized, our assumption should be reasonable. 

In general, however, the coefficients $\alpha_{j,i}$ are complex numbers. This means that in order to fully control the states $|\Psi_{i}\rangle$ one must not only control the absolute values of  $\alpha_{j,i}$ but also their phases. The later is extremely challenging; such degree of control is not present in current experiments. Thus, we have to assume that $\alpha_{j,i}$ acquire essentially random phases from pulse to pulse. 

Because the states $|\Psi_{i}\rangle$ will all have different relative phases for each laser pulse $i$, no pure state can be assigned to our source. Instead the state of the source is described by a density matrix $\rho_{s}$,
\begin{equation}
\rho_{s} = \frac{1}{M}\sum^{M}_{i=1}|\Psi_{i}\rangle\langle\Psi_{i}|.
\end{equation}
Expanding this sum using the definition of $|\Psi_{i}\rangle$ in Eq.(\ref{psi_i}) we obtain,
\begin{equation}\label{dm_discuss}
\rho_{s} = \sum^{N_{max}}_{j=0}p_{j}\rho_{j} +cross\:terms,
\end{equation}
here $p_{j}=|\alpha_{j}|^2, j=0,N_{max}$ is the probability to emit exactly $j$ electrons (note that by our assumption these probabilities are the same for every pulse $i$ so we have omitted the index $i$ all together), $\rho_{j}=|j^{el}\rangle\langle j^{el}|$  is a $j$-electron density matrix and its detailed structure will be discussed shortly. The cross terms in Eq.(\ref{dm_discuss}) are of the form $\rho_{mn}=|m^{el}\rangle\langle n^{el}|, m,n=0,\cdots,N_{max}$ where $|m^{el}\rangle$ and $|n^{el}\rangle$ are $m$-($n$-)electron states respectively. Each cross term  $\rho_{mn}$ comes with a pre factor 
\begin{equation}
p_{mn}=\frac{\sqrt{p_{m}p_{n}}}{M}\sum^{M}_{l=1}e^{i\phi_l(m,n)},
\end{equation}
where $p_{m}$($p_{n}$) are $m$-($n$-)electron generation probabilities, $\phi_l(m,n)$ is the random relative phase produced after the $l$-th laser pulse. Assuming that the phases $\phi_l(m,n)$ are distributed uniformly over $[0,2\pi]$ interval and $M\rightarrow\infty$ we conclude, using the central limit theorem, that $p_{mn}\rightarrow0$. In other words, the quantum coherence terms between various electron states vanish due to our inability to control their relative phases between pulses.

Taking the above discussion into account we can write the density matrix $\rho_{s}$ as
\begin{equation}\label{dm_s}
\rho_{s} =  \sum^{N_{max}}_{j=0}p_{j}\rho_{j}. 
\end{equation}
Let us now discuss the structure of $j$-electron states in more details. The simplest case is $j=0$, i.e. the vacuum. In this case $\rho_{0}=|0\rangle\langle 0|$. Let us further assume, since $N_{max}$ can be controlled experimentally e.g. by tuning the intensity of the laser, that $N_{max}=2$. This will simplify our calculations on one hand, yet, on the other, it will still provide us with an insight into the multi-electron quantum phenomenon.
  
\subsection{One-electron States}
Formally, we can write the one-electron density matrix $\rho_{1}=|1^{el}\rangle\langle1^{el}|$. But how does one model $|1^{el}\rangle$? Imagine for a second that we have chosen $|1^{el}\rangle=|1_{\bold{k}}\rangle$. In other words, we have assigned an eigenstate $|\bold{k}\rangle$ of the one-electron momentum operator to our particle. If we now measure the momentum of the electron we should get the value of $\bold{k}=(k_{x},k_{y},k_{z})$ as an outcome. Also, the uncertainty $(\Delta k_{x},\Delta k_{y},\Delta k_{z})$ of the measurement should be zero. On the other hand, the emission tip occupies a finite volume $\Delta V=\Delta x\Delta y\Delta z$. This provides limitations on the degree of spatial localization of the electron. Thus, using the Heisenberg uncertainty principle, we conclude that the uncertainty in measuring the momentum of the electron $\Delta k_{x}\Delta k_{y}\Delta k_{z}\ge\frac{1}{\Delta V}$. We immediately  see that our initial state assignment  $|1^{el}\rangle=|1_{\bold{k}}\rangle$ contradicts the uncertainty condition. Therefore, we should  rather treat the emitted electron as a wave packet i.e. as a superposition of different momentum eigenstates: 
\begin{equation}\label{1_el_state}
|1^{el}\rangle = \int^{+\infty}_{-\infty}d\bold{k} C(\bold{k})|1_{\bold{k}}\rangle,
\end{equation}
where $|C(\bold{k})|^2$ is a probability density function that describes particular details of the momentum distribution. Experimental data~\cite{el_spect} suggest that  $|C(\bold{k})|^2$ resembles a normal distribution with the mean $\bold{k_{0}}$,
\begin{equation}\label{1_el_pdf}
C(\bold{k})=[\frac{1}{2\pi\Delta k^{2}}]^{3/4}e^{-(\bold{k}-\bold{k_{0}})^{T}\Sigma^{-1}(\bold{k}-\bold{k_{0}})},
\end{equation}
where $\Sigma^{-1}=\frac{1}{4\Delta k^2}diag[1,1,1]$ and we assume that the variances $\Delta k_{x},\Delta k_{y},\Delta k_{z}$ are the same and equal to $\Delta k$.

\subsection{Two-electron States}

In quantum theory identical particles are assumed to be indistinguishable. This imposes a symmetry, under the operation of permutation, onto systems of two or more electrons. More specifically, the indistinguishability of electrons requires a multi-electron wave function to be antisymmetric under particle permutations. This leads to the Pauli exclusion principle (PEP) that forbids two electrons to be in the same one particle state. Therefore, the PEP should be take into account when constructing two-electron states.  

So far we have been ignoring spin degrees of freedom that electrons inherently have, concentrating only on the spatial part of electron states. However, the total state of an electron $|\Psi_{T}\rangle$ is a combination of the spatial and spin states, i.e. $|\Psi_{T}\rangle=|\psi_{spatial}\rangle\otimes|\phi_{spin}\rangle$. The PEP dictates that for two(or more) electrons the total electron state $|\Psi_{T}\rangle$ must be antisymmetric under particle permutations. Due to the direct product structure of $|\Psi_{T}\rangle$ only two possibilities are present: either $|\psi_{spatial}\rangle$ is antisymmetric and $|\phi_{spin}\rangle$ is symmetric or the other way around. We will talk about symmetric and antisymmetric spatial states in more details in a moment. First, let us concentrate on the spin part of $|\Psi_{T}\rangle$. 

It is well known that, there are three symmetric and one antisymmetric(think of the triplet and singlet states) spin states for two-electron systems. So, in general, $|\Psi_{T}\rangle$ can be decomposed into 
\begin{equation}\label{2el_decomp}
|\Psi_{T}\rangle = |\psi^{AS}\rangle\otimes\sum^{3}_{i=1}c_{i}|\phi^{S}_{i}\rangle+c_{4}|\psi^{S}\rangle\otimes |\phi^{AS}\rangle,
\end{equation}
where $|\phi^{S}_{i}\rangle$ denote the spin(symmetric) triplet states, $|\phi^{AS}\rangle$ is the spin singlet state, and we do not specify the explicit structure of the symmetric(antisymmetric) spatial components  $|\psi^{S}\rangle$($|\psi^{AS}\rangle$). In reality, no pure state like in Eq.(\ref{2el_decomp}) can be assigned to a nanotip electron emission source, because experimentally there is no control of spin polarization of emitted electrons. Assuming that all 4 spin states have equal generation probability(a fully unpolarized spin state), one can write down a density matrix describing the total two-electron spatial plus spin state:
\begin{equation}\label{2el_rho_ss}
\rho_{T} = \frac{3}{4}\rho_{AS}\otimes\sum^{3}_{i=1}|\phi^{S}_{i}\rangle \langle \phi^{S}_{i}|+\frac{1}{4}\rho_{S}\otimes |\phi^{AS}\rangle\langle \phi^{AS}|,
\end{equation}  
where $\rho_{AS}=|\psi^{AS}\rangle\langle \psi^{AS}|$ and $\rho_{S}=|\psi^{S}\rangle\langle \psi^{S}|$ are the spatial components of the total density matrix $\rho_{T}$. Moreover, since the spin degrees of freedom are neither measured nor controlled experimentally one can trace them out when describes experimental observations. This leads to the following two-electron density matrix,
\begin{equation}\label{2el_rho_ss_final}
\rho_{2} =  Tr_{spin} (\rho_{T}) =\frac{3}{4}\rho_{AS}+\frac{1}{4}\rho_{S}.
\end{equation}  

Now let us discuss the choice of $\rho_{S}$ and $\rho_{AS}$. We introduce two one-electron states $|\psi\rangle$ and $|\phi\rangle$ in the form of wave packets given in Eq.(\ref{1_el_state}). We assume that the wave packets are centered around $\bold{k_{0}}$ and $\bold{k^{\prime}_{0}}$ ($\bold{k_{0}}\ne\bold{k^{\prime}_{0}}$) in momentum space and have identical variances $\Delta {\bold k}$. Then define spatially symmetric and antisymmetric two-electron states $|2^{el}_{S}\rangle$ and $|2^{el}_{AS}\rangle$ as follows,
\begin{eqnarray}\label{2_el_state_s}
\!\! |2^{el}_{S}\rangle & = & \frac{1}{\sqrt{2(1+|\langle\psi|\phi\rangle|^{2})}}(|\psi\rangle\otimes|\phi\rangle + |\phi\rangle\otimes|\psi\rangle), \\
\label{2_el_state_as}
\!\!\! |2^{el}_{AS}\rangle & = & \frac{1}{\sqrt{2(1-|\langle\psi|\phi\rangle|^{2})}}(|\psi\rangle\otimes|\phi\rangle - |\phi\rangle\otimes|\psi\rangle),
\end{eqnarray}
where the overlap between the states $|\psi\rangle$ and $|\phi\rangle$ is given by,
\begin{equation}\label{2_el_olap}
|\langle\psi|\phi\rangle| = e^{-\frac{|\bold{k_{0}}-\bold{k^{\prime}_{0}}|^{2}}{8\Delta k^{2}}}.
\end{equation}
Substituting definitions of the states $|\psi\rangle$ and $|\phi\rangle$(Eqs.(\ref{1_el_state}-\ref{1_el_pdf})) into Eq.(\ref{2_el_state_s}-\ref{2_el_state_as}) we arrive at,
\begin{equation}\label{2_el_state_f}
|2^{el}_{S/AS}\rangle = N_{\pm}\cdot\int\!\!\!\!\int\limits^{+\infty}_{-\infty}d\bold{k_{1}}d\bold{k_{2}}\mathbf{C}_{\pm}(\bold{k_{1}},\bold{k_{2}}) |1_{\bold{k_{1}}} , 1_{\bold{k_{2}}} \rangle,
\end{equation} 
here we have introduced the functions
\begin{equation}
\mathbf{C}_{\pm} = \frac{e^{-\frac{|\bold{k_{0}}-\bold{k_{1}}|^{2}}{4\Delta k^{2}}}e^{-\frac{|\bold{k^{\prime}_{0}}-\bold{k_{2}}|^{2}}{4\Delta k^{2}}}\pm e^{-\frac{|\bold{k_{0}}-\bold{k_{2}}|^{2}}{4\Delta k^{2}}}e^{-\frac{|\bold{k^{\prime}_{0}}-\bold{k_{1}}|^{2}}{4\Delta k^{2}}}}{(2\pi\Delta k^{2})^{\frac{3}{4}}},
\end{equation} 
where $\mathbf{C}_{+}$ is a symmetric function($\mathbf{C}_{+}(\bold{k_{1}},\bold{k_{2}})=\mathbf{C}_{+}(\bold{k_{2}},\bold{k_{1}})$) corresponding to the state $|2^{el}_{S}\rangle$, $\mathbf{C}_{-}$ is an antisymmetric function($\mathbf{C}_{-}(\bold{k_{1}},\bold{k_{2}})=-\mathbf{C}_{-}(\bold{k_{2}},\bold{k_{1}})$) corresponding to the state $|2^{el}_{AS}\rangle$, and,
\begin{equation}
N_{\pm}= \frac{1}{\sqrt{2(1\pm|\langle\psi|\phi\rangle|^{2})}}.
\end{equation} 
The two-electron density matrix can be finally defined by substituting the states $|2^{el}_{AS}\rangle$ and $|2^{el}_{S}\rangle$ into Eq.(\ref{2el_rho_ss_final}).

\subsection{Electron Statistics}

Electron fields can be interrogated directly, using a particle detector that counts the number of electrons in a detection volume. A single electron detector provides information about electron densities at the location of the detector. One can also use multiple detectors in different locations to study temporal and spatial correlations  of electron densities. These data, in principle, is enough to gain a complete knowledge of the electron field of interest. 

The particle density operators at the position $\bold{x}$ is defined as $\hat{I}(\bold{x})=\hat{\Psi}^{\dagger}(\bold{x})\hat{\Psi}(\bold{x})$. Provided that we know the state of the field $\rho_{s}$, we can calculate the number of electrons at $\bold{x}$ as $n(\bold{x})=Tr(\hat{I}(\bold{x})\rho_s)$. In a similar fashion the number of electrons in a volume $\Delta V$ centered around $\bold{x}_{0}$ can be calculated as,
\begin{equation} \label{density}
n(\Delta V)=Tr\{\rho_s\int_{\Delta V} d^{3}x\hat{\Psi}^{\dagger}(\bold{x})\hat{\Psi}(\bold{x})\}.
\end{equation}

We recall that an important characteristics of quantum particle sources (electron, photon, etc) called the {\it degeneracy}~\cite{degeneracy} is defined as a number of particles per phase-space-cell volume. The volume of a phase-space cell is defined in terms of the Heisenberg uncertainty $\Delta\bold{x}\Delta\bold{k}=1$. Thus, using Eq.(\ref{density}), the electron source degeneracy in the vicinity of a point $\bold{x}_{0}$ reads,
\begin{equation} \label{degeneracy}
\delta=Tr\{\rho_s\int_{\Delta\bold{x}} d^{3}x\hat{\Psi}^{\dagger}(\bold{x})\hat{\Psi}(\bold{x})\},
\end{equation}
where $\Delta\bold{x}=\frac{1}{\Delta\bold{k}}$ is the position uncertainty calculated at the source($\bold{x}_{0}=0$). As a consequence of this definition, the electron degeneracy is the largest at the source and drops off as one moves away from it. This is true for both pulsed and continuous electron sources and is due to the dispersion of electron wave packets as they travel in space. Indeed, the largest contribution to the degeneracy at $\bold{x}_{0}$, according to Eq.(\ref{degeneracy}), comes from electron wave packets that are centered at $\bold{x}_{0}$. But their position uncertainty at $\bold{x}_{0}$ will be greater than it was at the source ($\Delta\bold{x}$). The number of electrons in the volume $\Delta\bold{x}$ at $\bold{x}_{0}$ will appear to be smaller than it was in the same volume at the source and, thus, the degeneracy will be less than it was at the source.  

The degeneracy attains its largest value when there is exactly one electron per phase-space cell (remember, electrons are fermions!). This would correspond to the degeneracy of 1. In this regime the quantum nature of electrons is the most pronounced. In reality, the best reported value of the degeneracy to date was achieved using a {\it continuous} electron emission source and is on the order of $10^{-4}$ ~\cite{Hasselbach}. This value is measured at the source and even though it is very small it allowed the experimental demonstration of the Hanbury-Brown Twiss effect for electrons in vacuum. It has also been argued that with the degeneracy values of $10^{-4}$ one could directly observe quantum behavior of electrons such as antibunching~\cite{Silverman}.
 
\section{Results: Pulsed Electron Source Degeneracy}\label{sec:results}

Calculating the electron degeneracy is now a straightforward task. We just need to combine Eq.(\ref{dm_s}) with Eq.(\ref{degeneracy}). Since the vacuum component of $\rho_{s}$ does not contribute to the degeneracy, we obtain,
\begin{equation}\label{d_total}
\delta=p_{1}\delta_{1}+p_{2}\delta_{2},
\end{equation}
where $p_{1},p_{2}$ are one-(two-)electron state probabilities, and $\delta_{1},\delta_{2}$ are correspondent degeneracy contributions. Note that the two-electron degeneracy $\delta_{2}$ contains contributions from spatially symmetric and antisymmetric two-electron states according to Eq.(\ref{2el_rho_ss_final}).  

So far we have been implicitly assuming that our electron source is stationary. In other words, electron statistics of the source, including the degeneracy, is invariant under time translation. This would apply well to electron sources operating in a continuous regime. However, the source we are interested in here is pulsed and the time interval between two consecutive electron emissions is several orders of magnitude larger than the duration of each pulse. In this situation, the value of the degeneracy will depend not only on the position in space but also time. Thus it is imperative that we account for time evolution when we calculate the degeneracy. 

\subsection{Degeneracy Calculation for Non-interacting Particles}\label{sec:nonintpart}
To include time evolution into our consideration, we need to either specify the time dependence of the electron field operator $\hat{\Psi}^{\dagger}(\bold{x},t),\hat{\Psi}(\bold{x},t)$ and keep the state $\rho_{s}$ in Eq.(\ref{dm_s}) time independent or keep the field operators time independent and instead use the time dependent density matrix $\rho_{s}(t)$. Here we choose the former approach. 

First, consider non-interacting electrons\footnote{We discuss effects of Coulomb interactions in Section \ref{sec:coulomb}}. The Hamiltonian describing the source is then given by,
\begin{equation}
\mathcal{H}=\int^{+\infty}_{-\infty}d\bold{k} \epsilon(\bold{k})c^{\dagger}_{\bold{k}}c_{\bold{k}},
\end{equation}
where $ \epsilon(\bold{k})=\frac{\hbar^{2}\bold{k}^2}{2m_{e}}$. Define the time-dependent electron field operators as follows,
\begin{eqnarray}\label{psihat}
\hat{\Psi}(\bold{x},t) & = & e^{-\frac{i}{\hbar}\mathcal{H}t}\hat{\Psi}(\bold{x},0)e^{\frac{i}{\hbar}\mathcal{H}t} \nonumber\\
						   & = & \frac{1}{\sqrt V}\int^{+\infty}_{-\infty}d\bold{k} e^{i \bold{k}\cdot \bold{x}}e^{-\frac{i}{\hbar}\epsilon(\bold{k})t} c_{\bold{k}} \nonumber\\
\hat{\Psi}^{\dagger}(\bold{x},t) & = &  \frac{1}{\sqrt V}\int^{+\infty}_{-\infty}d\bold{k} e^{-i \bold{k}\cdot \bold{x}}e^{\frac{i}{\hbar}\epsilon(\bold{k})t} c^{\dagger}_{\bold{k}}.
\end{eqnarray}
One last simplification is made before we calculate the time-dependent version of the degeneracy. Let us do the calculations for a one-dimensional case first. The generalization to three dimensions is then straightforward. However, a one-dimensional solution is algebraically tractable and thus easier to interpret. The one-dimensional one-particle component of the degeneracy $\delta_{1}$ reads,
\begin{eqnarray}\label{1el_degeneracy}
\delta_{1}(t) & = & Tr\{|1_{el}\rangle\langle 1_{el}|\int_{\Delta x} dx\hat{\Psi}^{\dagger}(x,t)\hat{\Psi}(x,t)\} \nonumber\\
				& = & \frac{1}{2}\{erf(\sqrt{2\alpha(t)}\Delta k(x_{2}-\frac{\hbar k t}{m_{e}}))\nonumber \\
				&    & -erf(\sqrt{2\alpha(t)}\Delta k(x_{1}-\frac{\hbar k t}{m_{e}}))\},
\end{eqnarray}
where $erf$ is the error function, $x_{1}=x_{0}-\frac{\Delta x}{2}$ and $x_{2}=x_{0}+\frac{\Delta x}{2}$(here $x_{0}$ and $\Delta x$  denote the location of the detector and its volume), and $\alpha(t)=\frac{m^{2}_{e}}{m^{2}_{e}+4\hbar^{2}\Delta k^{4}t^2}$ defines the spatial expansion of a Gaussian wave packet in time($\Delta x(t)=\frac{\Delta x(0)}{\sqrt{\alpha(t)}}$)~\cite{qmbook}. Note that the largest value of the degeneracy is given by $erf(\frac{1}{\sqrt{2}})<1$ and it is attained when $t=0$. This is because the one-electron state $|1_{el}\rangle$ at $t=0$ is a Gaussian wave packet with the position uncertainty $\Delta x = \frac{1}{\Delta k}$. Therefore, only a part of the wave packet will be detected in the volume $\Delta x$ around $x=0$. 

The result can be easily generalized to three dimensions, provided that there are no cross correlations between $k_{x},k_{y},k_{z}$. In this case the total one-particle contribution will be of the form $\delta^{3D}_{1}(t)=\delta^{x}_{1}(t)\delta^{y}_{1}(t)\delta^{z}_{1}(t)$, where $\delta^{i}_{1}(t),i=x,y,z$ is given in Eq.(\ref{1el_degeneracy}).
  
In a similar fashion, using the definition of the two-electron density matrix in Eq.(\ref{2el_rho_ss_final}), we can calculate the symmetric and antisymmetric components of the two-electron degeneracy in one dimension. If we define 
\begin{eqnarray}\label{definitions}
\xi_{1}(x,t,k) & = & erf(\sqrt{2\alpha(t)}\Delta k(x-\frac{\hbar k t}{m_{e}})),\nonumber \\
\xi_{2}(x,t,k_{1},k_{2}) & = & erf\{\sqrt{2\alpha(t)}\Delta k(x-\frac{\hbar (k_{1}+k_{2}) t}{2m_{e}}) \nonumber \\
							  &    & -\frac{i\sqrt{\alpha(t)}}{2\sqrt{2}\Delta k}(k_{1}-k_{2})\}, \nonumber \\
\theta_{2}(x,t,k_{1},k_{2}) & = & erf\{\sqrt{2\alpha(t)}\Delta k(x-\frac{\hbar (k_{1}+k_{2}) t}{2m_{e}}) \nonumber \\
							 &    & +\frac{i\sqrt{\alpha(t)}}{2\sqrt{2}\Delta k}(k_{1}-k_{2})\},
\end{eqnarray}  
then the two-particle contributions to the degeneracy read, 
\begin{eqnarray}\label{2el_degeneracy}
\delta_{2(S/AS)}(t) & = &  \frac{1}{2(1\pm\langle\psi|\phi\rangle^{2})}\{ \xi_{1}(x_{2},t,k_{0})-\xi_{1}(x_{1},t,k_{0})   \nonumber \\
				&  +  &  \xi_{1}(x_{2},t,k^{\prime}_{0})-\xi_{1}(x_{1},t,k^{\prime}_{0})  \nonumber \\
				& \pm & \langle\psi|\phi\rangle^{2}[ \xi_{2}(x_{2},t,k_{0},k^{\prime}_{0}) -  \xi_{2}(x_{1},t,k_{0},k^{\prime}_{0})  \nonumber \\
				&  +   &  \theta_{2}(x_{2},t,k_{0},k^{\prime}_{0}) -  \theta_{2}(x_{1},t,k_{0},k^{\prime}_{0})]\},
\end{eqnarray}
where the wave packet overlap $\langle\psi|\phi\rangle$ is defined in Eq.(\ref{2_el_olap}) and the plus(minus) sign corresponds to the contribution $\delta_{2S}(t)$($\delta_{2AS}(t)$) from the symmetric(antisymmetric) state $|2^{el}_{S}\rangle$($|2^{el}_{AS}\rangle$) in Eq.(\ref{2_el_state_f}). Let us examine the expression for $\delta_{2}(t)$ carefully. We notice that the difference $ \xi_{1}(x_{2},t,k_{0})-\xi_{1}(x_{1},t,k_{0}) = 2\delta_{1}(t,k_{0})$ can be interpreted as the "one-particle" degeneracy $\delta_{1}$ for the electron with the mean momentum $\hbar k_{0}$(compare to Eq.(\ref{1el_degeneracy})). Similarly, the difference $\xi_{1}(x_{2},t,k^{\prime}_{0})-\xi_{1}(x_{1},t,k^{\prime}_{0})=2\delta_{1}(t,k^{\prime}_{0})$ is the one-electron degeneracy for the electron the mean momentum $\hbar k^{\prime}_{0}$. The term in the square brackets in Eq.(\ref{2el_degeneracy}) contributes to the degeneracy only if the overlap between two electron wave packets is non zero. We call it the interference term. If the electrons are far apart in the momentum space, i.e. $|\hbar k_{0}- \hbar k^{\prime}_{0}|\gg 1$ and $\Delta k$ then they can occupy simultaneously the same positions in space and behave like classical particles. In this case the two-electron degeneracy reduces to a sum of one particle degeneracies and no quantum interference occurs. Putting all this together we rewrite Eq.(\ref{2el_degeneracy}) as follows,
\begin{eqnarray}\label{2el_degeneracy_f}
\delta_{2(S/AS)}(t) & = &  \frac{1}{(1\pm\langle\psi|\phi\rangle^{2})}\{ \delta_{1}(t,k_{0}) + \delta_{1}(t,k^{\prime}_{0}) \\
				&  \pm  & \langle\psi|\phi\rangle^{2}[ \delta_{\theta}(t)+\delta_{\xi}(t) ] \}\nonumber,\nonumber
\end{eqnarray}    
where we have defined
\begin{eqnarray}
 2\delta_{\theta}(t) & = & \theta_{2}(x_{2},t,k_{0},k^{\prime}_{0})- \theta_{2}(x_{1},t,k_{0},k^{\prime}_{0}) \\
 2\delta_{\xi}(t) & = & \xi_{2}(x_{2},t,k_{0},k^{\prime}_{0})- \xi_{2}(x_{1},t,k_{0},k^{\prime}_{0}).
\end{eqnarray}
The generalization of Eq.(\ref{2el_degeneracy_f}) to three dimension is also straightforward,
\begin{eqnarray}\label{2el_degeneracy_3D}
\delta^{3D}_{2(S/AS)}(t) & = &  \frac{1}{(1\pm\langle\psi|\phi\rangle^{2})}\{ \delta^{x}_{1}(t,k_{0_{x}}) \delta^{y}_{1}(t,k_{0_{y}}) \delta^{z}_{1}(t,k_{0_{z}}) \nonumber\\
				&  +  &  \delta^{x}_{1}(t,k^{\prime}_{0_{x}})\delta^{y}_{1}(t,k^{\prime}_{0_{y}})\delta^{z}_{1}(t,k^{\prime}_{0_{z}}) \\ 
				& \pm & \langle\psi|\phi\rangle^{2}[ \delta^{x}_{\theta}(t)\delta^{y}_{\theta}(t)\delta^{z}_{\theta}(t)+\delta^{x}_{\xi}(t)\delta^{y}_{\xi}(t)\delta^{z}_{\xi}(t) ]\}.\nonumber
\end{eqnarray}    
Finally, the total two-electron contribution to the degeneracy for our pulsed electron source reads,
\begin{equation}\label{2el_degeneracy_3D_tot}
\delta^{3D}_{2}(t)  = \frac{1}{4}\cdot \delta^{3D}_{2 S}(t) + \frac{3}{4}\cdot \delta^{3D}_{2 AS}(t).
\end{equation}    

\subsection{Results}\label{sec:results}

Let us now calculate the actual value of the degeneracy using realistic numbers for the source parameters. For that the following information is needed: 
\begin{itemize}
\item the position of the detector with respect to the source
\item the direction and the magnitude of the mean electron momentum (for both one- and two-electron states)
\item the momentum uncertainty in $x,y,z$ directions .
\end{itemize}
\begin{figure}[t]
\begin{center}
\includegraphics[scale=0.35]{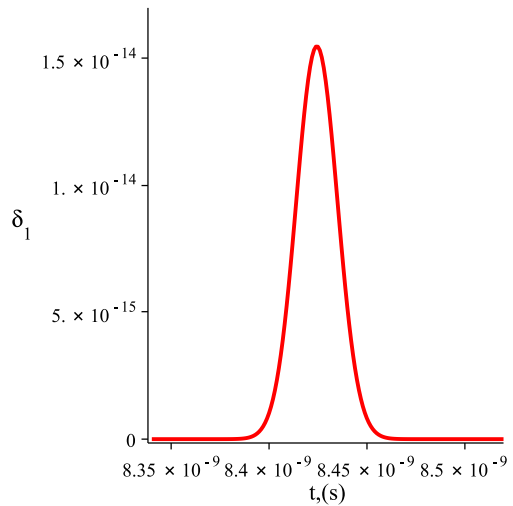}\vspace*{-0.5cm}
\end{center}
\caption{The one-electron degeneracy as a function of time. The position of the detector in meters is given by $\vec{\bold{r}}=(0.1,0,0)$. The mean value of the electron energy is taken to be $E_{0}=400 eV$ ($k_{0_{x}}=1.024\cdot10^{11}m^{-1},k_{0_{y}}=0,k_{0_{z}}=0$) and the energy spread $\Delta E=1eV$($\Delta k=1.28\cdot 10^{8}m^{-1}$ is assumed to be the same for $k_{x},k_{y},k_{z}$).}    
\label{fig:1eldegeneracy}
\end{figure}

\begin{figure}[t]
\begin{center}
\includegraphics[scale=0.35]{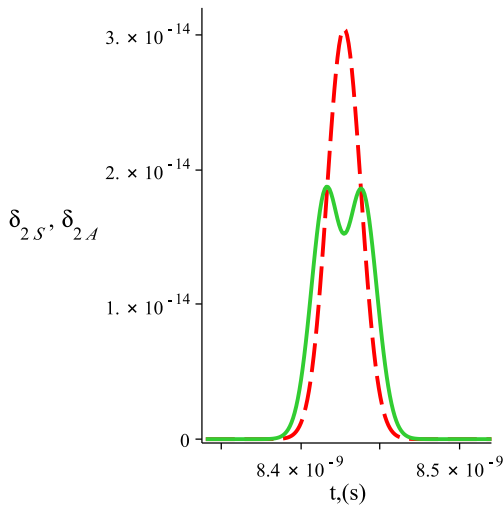}
\end{center}
\caption{The symmetric(dashed line) and antisymmetric(solid line) components of the two-electron degeneracy as a function of time Eq.(\ref{2el_degeneracy_3D}). The position of the detector in meters is given by $\vec{\bold{r}}=(0.1,0,0)$. The $x,y,z$ components of the mean momentum vector of the first electron are $k_{0_{x}}=1.024\cdot10^{11}m^{-1},k_{0_{y}}=0,k_{0_{z}}=0$ and the momentum spread $\Delta k=1.28\cdot 10^{8}m^{-1}$ is the same in all directions. For the second electron $k^{\prime}_{0_{x}}=k_{0_{x}}-0.5\cdot\Delta k,k^{\prime}_{0_{y}}=0,k^{\prime}_{0_{z}}=0$ and we also assume the same momentum spread $\Delta k=1.28\cdot 10^{8}m^{-1}$ in $x,y,$ and $z$ direction.}     
\label{fig:2eldegeneracy_SAS}
\end{figure}

\begin{figure}[t]
\begin{center}
\includegraphics[scale=0.35]{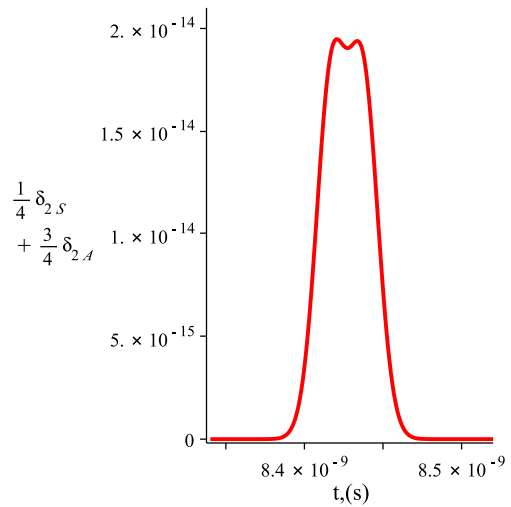}
\end{center}
\caption{The total two-electron degeneracy as a function of time defined in Eq.(\ref{2el_degeneracy_3D_tot}). The position of the detector, and all other relevant parameters are provided in the caption of Fig.(\ref{fig:2eldegeneracy_SAS}).}    
\label{fig:2eldegeneracy}
\end{figure}

\begin{figure}[t!]
\begin{center}
\includegraphics[scale=0.35]{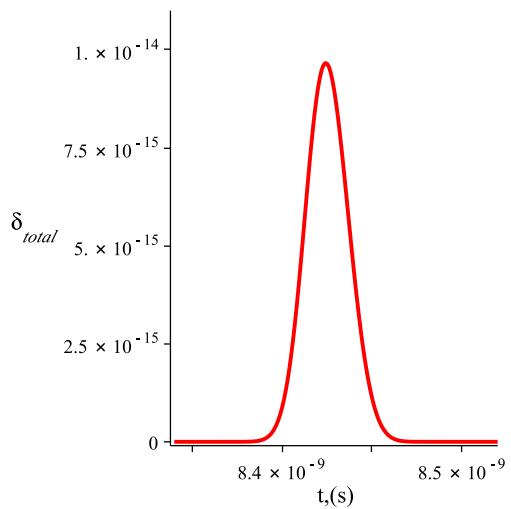}
\end{center}
\caption{The total (combined) degeneracy of the electron source as a function of time Eq.(\ref{d_total}). The values of the relevant parameters are the same as in Fig.(\ref{fig:1eldegeneracy}) and Fig.(\ref{fig:2eldegeneracy}). We set the one-electron probability $p_{1}=0.5$, and the two-electron probability $p_{2}=0.1$. }
\label{fig:totdegeneracy}
\end{figure}

We assume that the center of  the detector is positioned on the $x$ axis $0.1m$ away from the source. We also assume that the mean electron momenta for one- and two electron states have only non-zero components along the $x$ axis, i.e. $k_{0_{y}}=0,k_{0_{z}}=0$ and $k^{\prime}_{0_{y}}=0,k^{\prime}_{0_{z}}=0$. The experimental data~\cite{Bat1,sourcepar} suggests that the mean electron energy for a pulsed nanotip source $E_{0}$ can be, for example, $400 eV$  and $\Delta E\approx 1eV$. These translate into $k_{0_{x}}=1.024\cdot10^{11}m^{-1}$ and $\Delta k_{x,y,z}=1.28\cdot 10^{8}m^{-1}$. Using these values we plot the one-electron, two-electron, and total electron 3D degeneracy of the source as a function of time in Figs.(\ref{fig:1eldegeneracy}-\ref{fig:totdegeneracy}). Also, to compare the degeneracy of our source to the best available {\it continuous} sources we compute the value of the degeneracy at the source. For the above choice of parameters $\delta^{3D}_{tot}\approx 0.2$, i.e. three orders of magnitude larger than the value reported in~\cite{Hasselbach}. Larger values of the degeneracy for pulsed sources are expected and can be explained by  much higher electron current densities generated by pulsed laser light.

A discussion of the results is in order. First, let us examine the one-electron contribution to the degeneracy in Fig.(\ref{fig:1eldegeneracy}). As one can see the shape of the curve resembles a Gaussian. It is due to the fact that our one-electron state is a Gaussian wave packet traveling through the detection volume. When the center of the one-electron wave packet reaches the middle point of the detector($\vec{\bold{r}}=(0.1,0,0)$) the degeneracy attains its maximum value($\approx 1.55\cdot 10^{-14}$). The reason why this value is $\ll 1$ is because the wave packet spreads out in all spatial directions as a result of the propagation, however, our detection volume equals the initial spatial spread of the wave packet at $t=0$. Thus the value of the degeneracy depends on how fast the initial wave packet disperses and how far from the source the detector is.

As we have already discussed, the two-electron contribution to the degeneracy has two components stemming from the symmetrized and antisymmetrized states (see Fig.(\ref{fig:2eldegeneracy_SAS})). The symmetric contribution resembles a gaussian, meaning that both electrons can be detected at the center of the detector at the same time. This is not surprising since the underlying two-electron wave packet is symmetric, i.e. electrons behave like bosons despite the fact that they are fermions. On the other hand, the antisymmetric contribution to the degeneracy has two maxima that correspond to the passage of either one of the electrons through the detector. In between those maxima there a minimum that is attained when the center of the antisymmetrized  two-electron wave packet passes through the center point of the detector. Why is that? Recall that for the two-electron states we have set the value of $k^{\prime}_{0_{x}}=k_{0_{x}}-0.5\cdot\Delta k$. This means that the antisymmetric two-electron state consists of two, strongly overlapping(in momentum space), wave packets which, according to the PEP, may not overlap spatially. In other words, the probability to detect two electrons in the center of the wave packet drops down to zero. As the wave packet propagates the probability is still zero in the center but the wings of the wave packet will disperse spatially much like in the one-electron case. Thus the value of the antisymmetric component of the degeneracy will experience a reduction when the center of the two-electron packet is at the central point of the detection volume. 

The total two-electron degeneracy contribution is a weighted sum of the symmetric and antisymmetric components according to Eq.(\ref{2el_degeneracy_3D_tot}).  We plot it on Fig.(\ref{fig:2eldegeneracy}) as a function of time. Its distinct feature is a dip in the center of the gaussian that is attributed to the destructive interference of two electrons constituting the antisymmetric state $|2^{el}_{AS}\rangle$. To observe this two-electron dip experimentally one needs a detector with time resolution on the order of $20$ picoseconds which is a typical time resolution of top of the line detectors.
  
 The total electron degeneracy is a weighted sum of the one- and two-electron contributions. It is depicted in Fig.(\ref{fig:totdegeneracy}). The weighting coefficients, i.e. the probabilities of one- and two-electron states $p_{1}$ and $p_{2}$, are chosen such that the the one-electron events are dominant i.e. $p_{1}=0.5$, $p_{2}=0.1$. Therefore, it is not a surprise that the total degeneracy is close in its shape to the one-electron component on Fig(\ref{fig:1eldegeneracy}). This particular choice of $p_{1}$ and $p_{2}$ also implies a high probability of vacuum generation since $p_{0}=1-p_{1}-p_{2}$. In general, $p_{1}$ and $p_{2}$ can be controlled experimentally. For instance, by adjusting the power of laser pulses. Then, assuming that electron statistics is governed by the Poisson distribution, the two-electron events can be made dominant ($p_{2}\ge p_{1}$) and one could study electron-electron interference by just measuring the total degeneracy as a function of time. 
 
One final remark on how the degeneracy can be measured. Of course, in an experiment it is rather the brightness, defined as the number of particles per unit area per unit solid angle per second, that is measured routinely. Note, however, that the brightness is closely related to the degeneracy~\cite{degeneracy}, and, therefore the value of the degeneracy at the source can be determined by measuring the brightness.
   
\section{Quantum Interference Effects with Pulsed Electrons}\label{sec:interference}

Quantum interference between two electrons is present in the time domain. The consequence of this for degeneracy in the two-electron contribution was discussed in Section \ref{sec:results}. There, the fact that electrons obey the PEP and, therefore, may not be present at the same location simultaneously led to a dip in the degeneracy as a function of time(see Fig.(\ref{fig:2eldegeneracy})). The same effect leads to a spatial interference pattern for two electrons. To observe it, one can use two detectors and record the coincidence rate of the detectors as a function of their relative distance. The resulting spatial pattern will have a dip at the point where the relative distance is zero. Note that this measurement requires two detectors, whereas to observe the dip in the two-electron degeneracy only one detector is needed. In this section we discuss two-electron spatial interference effects that can be observed with one and two detectors and study how useful those can be for discriminating between states with different number of electrons. 

\subsection{Discrimination Between One- and Two-electron Components}

The degeneracy of an electron source can be measured directly, as discussed in Section~\ref{sec:results}. However, the value of the degeneracy does not provide complete information about the density matrix of the source. In general, to characterize the source one needs to perform a complete state tomography. Without any assumptions about the source, this may become a practically impossible task. On the other hand, if the experimentator could make some reasonable assumptions about the source, based on his/her knowledge of the setup, then the state tomography might not be needed. For example, with an assumption about the structure of the density matrix, such as the one in Eq.(\ref{dm_s}), one, in principle, can calculate the probabilities to detect vacuum, one-, and two-electron states(provided $N_{max}=2$) by fitting the plot of the total degeneracy in Fig.(\ref{fig:totdegeneracy}).      
 
In certain situations, for instance, when the time resolution of one's detector is not high enough, measuring the degeneracy as a function of time might not be practical. Still, one would like to find probabilities of different components of the density matrix describing the source. Here is how it can be done with just one detector.
  
First, the probability density to detect an electron, described by the state $\rho_{1}=|1^{el}\rangle \langle 1^{el}|$ in Eq.(\ref{1_el_state}), at a position $x$ reads,
\begin{eqnarray}\label{1_el_density}
P_{1}(x,t) & = & \langle 1^{el}|\hat{\Psi}^{\dagger}(x,t)|0\rangle\langle 0|\hat{\Psi}(x,t)|1^{el}\rangle \nonumber \\
& = & \frac{\Delta k \sqrt{2 \alpha(t)}}{\sqrt{\pi} }e^{-2\Delta k^2\alpha(t)(x-\hbar k_{0}t/m)^2} ,
\end{eqnarray}
where $\hat{\Psi}(x,t)$ was previously defined in Eq.(\ref{psihat}) and, as before, $\alpha(t)=\frac{m^{2}_{e}}{m^{2}_{e}+4\hbar^{2}\Delta k^{4}t^2}$. If we set time $t=0$ we immediately recover a familiar gaussian distribution $P_{1}(x)=\frac{\Delta k \sqrt{2}}{\sqrt{\pi} }exp(-2\Delta k^2x^2) $ centered around $x=0$ with the width $1/2\Delta k$. 

On the other hand, in the case of the two-electron state $\rho_{2}$ in Eq.(\ref{2el_rho_ss_final}), the joint probability to detect one electron in the position $x_{1}$ at $t=0$, and the other one in the position $x_{2}$ at $t=0$ for symmetric/antisymmetric component is given by,
\begin{eqnarray}
P_{S/AS}(x_{1},x_{2})  & = & |\langle 2^{el}_{S/AS}|\hat{\Psi}^{\dagger}(x_{1})\hat{\Psi}^{\dagger}(x_{2})|0\rangle|^{2} \nonumber \\ 
= \frac{2\Delta k^{2}e^{-2\Delta k^2(x^{2}_{1}+x^{2}_{2})}}{\pi (1\pm\langle \psi |\phi \rangle^{2})}& \times& (1\pm \cos{X(k_{0}-k^{\prime}_{0})} ), 
\end{eqnarray}
where $X=x_{1}-x_{2}$. The total joint probability $P(x_{1},x_{2})$ reads,
\begin{equation}\label{2_el_density}
P(x_{1},x_{2})=\frac{1}{4}P_{S}(x_{1},x_{2})+\frac{3}{4}P_{AS}(x_{1},x_{2}).
\end{equation}
 The probability $P(x_{1},x_{2})$ -- a two-detector quantity -- can be used to construct a one detector probability $P(x)$. For example, $P(x)$ can be the probability to detect an electron(does not matter which) at a position $x$, provided that the other electron is traced out. It can be obtained by integrating the probability $P(x_{1},x_{2})$ over either one of the variables i.e. $P(x)=\int_{-\infty}^{\infty}dy P(x,y) = \int_{-\infty}^{\infty}dz P(z,x)$. Physically, $P(x)$ can be measured with just one detector, by shutting it off once it has detected an electron(note that this is different from detecting both electrons at $x$; the probability for this event is given by $P(x,x)$). By integrating Eq.(\ref{2_el_density}) over one of the variables we obtain,
\begin{equation}\label{1_el_density_2el}
P(x) = \frac{1}{4}P_{S}(x)+\frac{3}{4}P_{AS}(x),
\end{equation} 
where the following quantities were introduced
\begin{eqnarray}\label{1_el_density_2el_SAS}
\!\! P_{S/AS}(x) & = & \frac{\sqrt{2}\Delta ke^{-2\Delta k^2x^{2}}}{\sqrt{\pi} (1\pm\langle \psi |\phi \rangle^{2})}(1\pm\langle \psi |\phi \rangle\cos{x(k_{0}-k^{\prime}_{0})}). \nonumber \\
\end{eqnarray}
Once again, Eq.(\ref{1_el_density_2el}) represents the probability density to detect just one of the two electrons that are described by the state $\rho_{2}$ while the destiny of the other electron is completely ignored. If the overlap between single-electron states $|\psi\rangle$ and $|\phi\rangle$ is non-negligible, then, by examining the Eq.(\ref{1_el_density_2el_SAS}), we conclude that $P(x)$ will be reduced at the center position ($x=0$) when compared to the probability $P_{1}(x)$ in Eq(\ref{1_el_density}). This is due to the interference terms containing $-\cos{x(k_{0}-k^{\prime}_{0})}$. The interference stems from the indistinguishability of electrons. When just one out of two electrons is detected, we cannot tell which electron was detected, and thus the interference occurs. 
\begin{figure}[th]
\begin{center}
\includegraphics[scale=0.35]{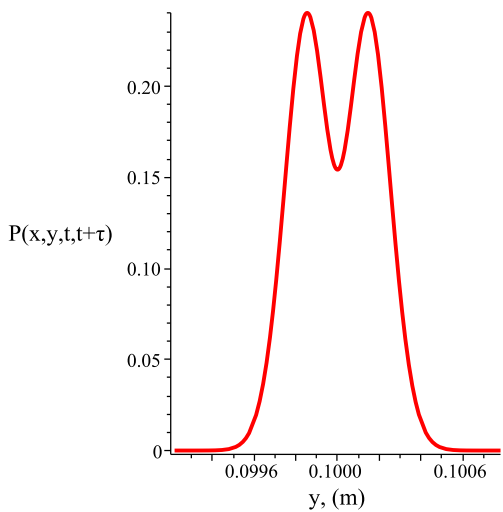}
\end{center}
\caption{Joint probability to detect one electron by a detector located at position $(x,t)$ and another electron by a detector at $(y,t+\tau)$ as a function of the position of the second detector. The first detector is fixed at $x=0.1 m$, $\tau=0$, and $t=8.427\cdot 10^{-9} s$. The parameters for the two-electron wave packet used in the simulations are $k_{0_{x}}=1.024\cdot10^{11}m^{-1},(k_{0_{y}},k_{0_{z}})=0$ for the first electron and $k^{\prime}_{0_{x}}=k_{0_{x}}-0.5\cdot\Delta k,(k^{\prime}_{0_{y}},k^{\prime}_{0_{z}})=0$ for the second electron. The momentum spread $\Delta k=1.28\cdot 10^{8}m^{-1}$ is the same for both electrons in all directions.}
\label{fig:hbt}
\end{figure}

The one-detector two-electron interference described here is a spatial version of the temporal behavior of the two-electron degeneracy component studied in Section~\ref{sec:results}. Both effects can be explained by the indistinguishability of individual electrons in a two-electron wave packet. These effects illustrate how two-electron phenomena are different from one-electron phenomena, and, thus, they can be naturally used to discriminate one- and two-electron contributions in the density matrix describing the source. 

Recall the density matrix $\rho_{s}$ describing the source and defined in Eq.(\ref{dm_s}). Imagine that from some experimental consideration we know that $N_{max}=2$, however, what is unknown are the probabilities of one- and two-electron outcomes $p_{1},p_{2}$. Moreover, we only have one electron detector that we can move around such that a spatial distribution of electron detection events can be recorded. Of course, measuring the probability of having no electrons $p_{0}$ is relatively easy since we just need to record no-click events and normalize their number to the total number of measurements. Thus, $p_{2}$ can be expressed as a linear function of $p_{1}$ i.e. $p_{2}=1-p_{0}-p_{1}$. Then, if we measure only one-electron events at different locations and shut the detector off right away after it clicks, the resulting probability of one-electron events is given by $p_{1}P_{1}(x)+p_{2}P(x)$, where $P_{1}(x)$ and $P(x)$ were defined in Eq.(\ref{1_el_density}) and Eq.(\ref{1_el_density_2el}) respectively. So, given the measured data, we can determine the value of $p_{1}$ and $p_{2}$ by fitting the data with the one-parameter function $p_{1}P_{1}(x)+p_{2}P(x)$, and, thus resolve the one- and two-electron components of the density matrix.

\subsection{Hanbury Brown and Twiss Effect for Two-electron States}

In many situations, one may be interested in coherence properties of the electron source (e.g. transverse (longitudinal) coherence, and (or) the effective size of the source) rather then in its density matrix. To determine those quantities, a two-detector measurement technique developed by Hanbury-Brown and Twiss (HBT) in stellar astronomy can be borrowed. 

In their groundbreaking work, HBT~\cite{HBT} were interested in two-point {\it intensity} correlations of the light coming from a distant star. Using this data they were able to measure the angular diameter of the distant star. The joint two-detector probability $P(x_{1},x_{2})$ given in Eq.(\ref{2_el_density}) is the equivalent of the HBT measurement for {\it  pulsed} electron sources at $t=0$. Naturally, a pair of spatially separated electron detectors is required to measure this probability. Note that Eq.(\ref{2_el_density}) describes a particular measurement setup where one is interested in detecting two electrons in two different positions at the same instance in time(namely at $t=0$). For electron sources whose electron statistics does not explicitly depend on time (e.g. continuous sources) Eq.(\ref{2_el_density}) also provide the probability of simultaneous detection of two electrons at any instance in time $t>0$, provided that electron dispersion effects are taken into account. For pulsed sources the expression in Eq.(\ref{2_el_density}) has to be modified to account for time dependence in two-electron correlations at different times (e.g. $x,t$ and $y,t+\tau$). For instance, in Fig.(\ref{fig:hbt}), the joint probability to detect one electron at ($x=0.1 m, t=8.427\cdot 10^{-9} s$) as a function of the position of the second detector at the same instance in time ($t=8.427\cdot 10^{-9} s$) is plotted. We observe that when the positions of both detectors coincide ($x=y=0.1 m$), the probability to find both electrons is reduced. This is a direct consequence of the PEP. The electrons that belong to the antisymmetric two-electron state are not allowed to be in the same position.  

\section{Effects of Coulomb Repulsion}\label{sec:coulomb}

So far no interaction between electrons has been assumed. In reality, electrons in a two-electron state like, for example, the one described in Eq.(\ref{2_el_state_f}) are close enough spatially to experience strong Coulomb repulsion. But will this affect electron-electron interference  and reduce the degeneracy? Intuitively it is clear that the relative momentum between two electrons will be increased by Coulomb repulsion, pushing them away from each other in momentum space. This, in turn, will result in a reduction of quantum interference. At first, one might, prematurely, conclude that all is lost and Coulomb repulsion would ruin quantum effects. However, if the rate of electron wave packet dispersion -- a purely quantum feature -- is larger than the rate of relative momentum change due to Coulomb repulsion then quantum effects will still dominate!
 
Let us first estimate the effect of Coulomb interaction and then compare it to the rate of wave-packet dispersion. For the former, consider two classical particles of equal charge $q_{e}$ and mass $m_{e}$. In the center-of-mass (COM) reference frame relevant system variables are the relative momentum $p$ and the relative position of the particles $x$. If the relative momentum of the particles at $t=0$ is $p_{0}$ and the relative distance is $x_{0}$ then, by employing the energy conservation argument, we arrive at the following relationship between $x$ and $p$ at $t>0$,
\begin{equation}
\frac{p^{2}}{m_{e}}+\frac{k_{e} q^{2}_{e}}{x}=E_{0}=\frac{p_{0}^{2}}{m_{e}}+\frac{k_{e} q^{2}_{e}}{x_{0}},
\end{equation} 
which leads to the expression for $p$ as a function of $x$,
\begin{equation}\label{eq:relmom}
p=\sqrt{m_{e}E_{0}-\frac{m_{e}k_{e} q^{2}_{e}}{x}}.
\end{equation}  
\begin{figure}[t]
\begin{center}
\includegraphics[scale=0.35]{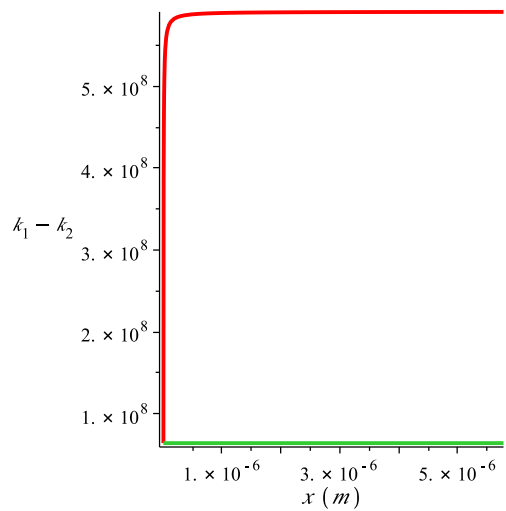}
\end{center}
\caption{Relative momentum as a function of relative coordinate for two classical electrons (red solid line). The initial relative momentum is represented by the green solid line parallel to the $x$ axes. }
\label{fig:coulomb}
\end{figure}
To proceed further, an estimate of the value of $E_{0}$ is needed. We will assume that $p_{0}=\hbar (k_{0_{x}}-k^{\prime}_{0_{x}})=0.5\cdot\hbar\Delta k$ and we will use $\Delta k=1.28\cdot 10^{8}m^{-1}$ -- the same values that we have used to describe two-electron states in the preceding section. The initial relative distance $x_{0}$ can then be estimated from the distance between the maxima of the two-electron probability density as $x_{0}=\frac{1}{\sqrt{2}\Delta k}=5.52\cdot10^{-9}m$. On Fig.(\ref{fig:coulomb}) the relative momentum normalized to $\hbar$ as a function of the relative distance starting from $x_0$ is plotted. As expected, the relative momentum increases with the relative distance until it reaches its maximum -- a point where Coulomb repulsion is negligible. Using Eq.(\ref{eq:relmom}) the terminal relative momentum $p_{t}=\sqrt{m_{e}E_{0}} $ when $x\rightarrow\infty$, and  the ratio $p_{t}/p_{0}$ can both be determined. For our choice of  parameters, this ratio is $\approx 9$. Next we calculate the relative distance $x_{det}$ between the electrons when their COM reaches the detector placed $0.1m$ away from the source by integrating Eq.(\ref{eq:relmom}). After some algebra we find that $x_{det}=5.77\cdot10^{-4} m$. Note that the relative momentum between electrons at  $x_{det}$ is very close to its terminal value(see Fig.(\ref{fig:coulomb})). Finally, the effect of Coulomb repulsion for two electrons can be estimated by the ratio $x_{0}/x_{det}\approx 0.95\cdot 10^{-5}$. 

On the other hand, the rate of quantum dispersion of a two-electron wave packet in one dimension is given by the ratio $\Delta x(0)/ \Delta x(t_{det})=\sqrt{\alpha(t_{det})}$, where $\alpha(t)$ was defined in Section \ref{sec:nonintpart}. Calculating it for the same parameter values as in the Coulomb repulsion problem gives $\Delta x(0)/ \Delta x(t_{det})\approx 3.1\cdot 10^{-5}$. Finally, by comparing magnitudes of $x_{det}$ and $\Delta x(t_{det})$ we conclude that for our parameters Coulomb repulsion will, indeed, dominate quantum effects by roughly one order of magnitude.  

Fortunately, there is a way to enhance quantum effects while keeping Coulomb repulsion nearly constant. Note that the strength of Coulomb repulsion depends on the initial distance between electrons. The further away the electrons are, the smaller the effect of Coulomb interaction is. But simply putting the electrons further apart without increasing their longitudinal coherence $\Delta x$ will again result in Coulomb domination.  That means that to make quantum effects dominant it is necessary to increase electron coherence. Here is how that can be done in practice. In reality, two-electron wave packets are three dimensional. They are characterized by the longitudinal coherence $\Delta x$ and transverse coherence $\Delta y\cdot \Delta z$. The longitudinal coherence is related to coherence time $\Delta t$  and is hard to control in an experiment. On the other hand, the transverse coherence can be easily improved by placing a transversal filter in combination with a quadrupole lens. Increasing transverse coherence, while keeping longitudinal coherence intact, will result in a wave packet shaped like a pancake where longitudinally the electrons are close, and transversally far from each other. 

Consider for now a two-dimensional wave packet with the initial longitudinal coherence $\Delta x = 1/\Delta k$, where $\Delta k$ is the same as in the one-dimensional calculations. Assume that the initial transverse coherence $\Delta y$ is ten times better than $\Delta x$, i.e. $\Delta y=10\cdot \Delta x$. Like in the one-dimensional case, quantum effects can be estimated by the ratio $\Delta x(0)\Delta y(0)/ \Delta x(t_{det})\Delta y(t_{det})$ which is $\approx 9.75\cdot10^{-8}$. The effect of Coulomb repulsion can be also estimated by upgrading the one-dimensional model to two dimensions. The initial distance between the electrons is $r_{0}=5.55\cdot10^{-8}m$. The relative distance change at $t=t_{det}$ due to Coulomb repulsion is $r^{2}_{0}/r^{2}_{det}\approx 9.35\cdot 10^{-8}$. And one immediately concludes that quantum effects are now of the same order as Coulomb repulsion. The quantum effects can further be made dominant by considering three-dimensional wave packets with improved transverse coherence.

\section{Summary and Conclusions}\label{sec:last}
We developed a formalism that can be applied to the description of both pulsed and continuous nano tip electron emission sources. To make quantitative predictions about particle statistics  of a pulsed electron source, a model, describing the density matrix of the source, was introduced. Based on this model, the quantum degeneracy of the source was calculated. The degeneracy at the source is predicted to be three orders of magnitude higher than the best previously reported value for continuous sources. Quantum effects due to electron-electron interference, that can be observed while measuring, for example, the degeneracy as a function of time were studied. It is argued that these effects can readily be observed experimentally. The relationship to other electron interference effects, such as the HBT effect, were discussed. The effect of Coulomb repulsion was estimated and it was shown that quantum effects can be made dominant over it. 

For the newest pulsed electron sources the expected vales of degeneracy are such that the Pauli Exclusion Principle should be considered to describe its properties correctly. Furthermore given the developments of temporal lenses to control and reverse electron dispersion, degeneracy is also expected to be essential at the image location of such temporal lenses. These developments appear to usher in a new field of electron quantum optics. Current efforts are underway in the laboratory~\cite{HB} to observe such effects, and it is our hope that the above formalism provides some guidance to this and other similar future experiments. 

\section{Acknowledgments}
H.B. gratefully acknowledges support by the National Science Foundation, Grant No. 2505210148001, P.L. acknowledges support from NSF MRSEC (Grant No. DMR-0820521) and Nebraska Research Initiative.


\begin{thebibliography}{99}
\bibitem{fnote1} 
The index $\bold{k}$ is used to catalog single-particle wave functions and its meaning depends on the context. Here, we refer to $\bold{k}$ as to an index describing the eigenstates of a one-electron momentum operator normalized to $\hbar$.
\bibitem{fnote2} 
At this point we do not specify the explicit structure of N-electron states. It can either be an N-electron wave packet or an eigenstate of the momentum operator.
\end{thebibliography}
\end{document}